\pdfoutput=1 
\documentclass[a4paper, 11pt]{article}
\usepackage[dvipsnames]{xcolor}

\usepackage{jheppub} 

\makeatletter  
\gdef\@fpheader{}  
\makeatother 

\usepackage{bm}
\usepackage{braket}

\usepackage{blindtext}
\usepackage{epsfig, cancel}

\usepackage{latexsym}
\usepackage{natbib, comment}
\usepackage{mathrsfs,amsmath,amssymb,amsthm,amsfonts,tikz,graphicx,accents,hyperref,color}
\usepackage{url}
\usepackage{dcolumn}
\usepackage{multirow}
\usepackage{color}
\usepackage{cancel}
\usepackage{soul}
\usepackage[normalem]{ulem}
\usepackage{txfonts}
\usepackage{epsfig}
\usepackage{psfrag}
\usepackage{subfigure}
\hypersetup{colorlinks=true}
\usepackage{mathtools}
\usepackage{enumitem}
\usepackage{float}
\usetikzlibrary{intersections,calc}

\def\H0{{\text{H}\hspace*{-2.05mm}\text{H} 0\hspace*{-1.35mm}0\ }}

\usepackage[all]{xy}

\usepackage{amssymb}
\usepackage{ragged2e}
\usepackage{slashed}
\usepackage{slashed,ccaption}
\usepackage{multirow}

\usepackage{caption}

\usepackage{array}
\hypersetup{ linktoc=all,
    colorlinks, linkcolor={palatinateblue},
    citecolor={brightpink}, urlcolor={amaranth}
}

\graphicspath{{Images/}}

\renewcommand{\d}[1]{\ensuremath{\operatorname{d}\!{#1}}}

\DeclareSymbolFont{extraup}{U}{zavm}{m}{n}
\DeclareMathSymbol{\varheart}{\mathalpha}{extraup}{86}
\DeclareMathSymbol{\vardiamond}{\mathalpha}{extraup}{87}
\makeatletter
\renewcommand*{\@fnsymbol}[1]{\ensuremath{\ifcase#1\or \clubsuit \or \vardiamond \or \varheart\or
    \spadesuit\or \mathparagraph\or \|\or **\or \dagger\dagger
    \or \ddagger\ddagger \else\@ctrerr\fi}}
\makeatother

\definecolor{rosy}{RGB}{230,235,252}
\definecolor{myframetitle}{RGB}{90,89,170}
\definecolor{myblocktitle}{RGB}{140,185,249}
\definecolor{mytitle}{RGB}{10,80,26}

\definecolor{darkgreen}{RGB}{27,130,45}
\definecolor{darkblue}{rgb}{0,0,0.3}
\definecolor{darkred}{rgb}{0.7,0,0}

\definecolor{light gray}{RGB}{220,220,220}
\definecolor{dark purple}{RGB}{108,0,217}
\definecolor{pink}{RGB}{190,20,100}
\definecolor{orang}{RGB}{193,63,0}
\definecolor{green}{RGB}{11,98,17}
\definecolor{darkpink}{RGB}{153,0,76}
\definecolor{bluegreen}{RGB}{0,102,102}
\definecolor{greenlagan}{RGB}{0,102,0}
\definecolor{redgreen}{RGB}{102,102,0}
\definecolor{Redgreen}{RGB}{153,76,0}
\definecolor{vividviolet}{rgb}{0.62, 0.0, 1.0}
\definecolor{amaranth}{rgb}{0.9, 0.17, 0.31}
\definecolor{palatinateblue}{rgb}{0.15, 0.23, 0.89}
\definecolor{brightpink}{rgb}{1.0, 0.0, 0.5}
\definecolor{cornflowerblue}{rgb}{0.39, 0.58, 0.93}
\definecolor{deepcarminepink}{rgb}{0.94, 0.19, 0.22}
\definecolor{radicalred}{rgb}{1.0, 0.21, 0.37}

\newcommand\snote[1]{\textcolor{darkpink}{\bf [Sh:\,#1]}}

\DeclareFontFamily{OT1}{rsfs}{}

\DeclareFontShape{OT1}{rsfs}{m}{n}{ <-7> rsfs5 <7-10> rsfs7 <10->rsfs10}{} 

\DeclareMathAlphabet{\mycal}{OT1}{rsfs}{m}{n}

\newcommand{\bcC}{\boldsymbol{\mathcal{C}}}

\newcommand{\bcD}{\boldsymbol{\mathcal{D}}}
\newcommand{\bcB}{\boldsymbol{\mathcal{B}}}
\newcommand{\bcN}{\boldsymbol{\mathcal{N}}}

\newcommand{\be}{\begin{equation}}
\newcommand{\ee}{\end{equation}}

\usepackage[most]{tcolorbox}

\tcbset{highlight math style={left=02mm,right=02mm,top=02mm,bottom=02mm}} 
\usepackage{empheq}

\newcommand\inbox[1]{\tcbset{fonttitle=\scriptsize} \tcboxmath[colback=white,colframe=black!70]{#1}}

\newcommand{\Ottbar}{{\mathcal{T}\hspace*{-1mm}{\mathcal{T}}}}

\newcommand{\ttbarb}{\mathrm{T}\hspace*{-1mm}\mathrm{T}}

\begin{document}




\title{Freelance Fluid/Gravity Correspondence, 3d Analysis}

\author[a]{M.M.~Sheikh-Jabbari}
\author[b,a]{, V.~Taghiloo}

\affiliation{$^a$ School of Physics, Institute for Research in Fundamental
Sciences (IPM),\\ P.O.Box 19395-5531, Tehran, Iran}
\affiliation{$^b$ Department of Physics, Institute for Advanced Studies in Basic Sciences (IASBS),\\ 
P.O. Box 45137-66731, Zanjan, Iran}
\emailAdd{
jabbari@theory.ipm.ac.ir, v.taghiloo@iasbs.ac.ir}
\abstract{Freelance holography program is an extension of gauge/gravity correspondence, where the gravity theory is defined on a portion of AdS with an arbitrary timelike boundary, with any desired boundary conditions. It is also known that gauge/gravity correspondence admits a fluid/gravity correspondence limit, where the gauge theory side is well described by a fluid. In this work, combining the two, we work through ``freelance fluid/gravity''. In particular, we study in detail the 2d fluid (3d Einstein gravity) case, where one has a good analytical control over the bulk equations due to their integrability and absence of viscosity in the 2d fluid. We study consistency and validity requirements for the freelance fluid/gravity and how the fluid changes along the renormalization group (RG) flow. We prove the $v_g$-theorem, stating that the group velocity of fluid waves $v_g$ is a decreasing function as we move toward the infrared region along the RG flow, regardless of the adopted boundary conditions. We also study examples of holographic fluid with various asymptotic boundary conditions.   }
\maketitle

\section{Introduction}\label{sec:Introduction}
The salient feature of gravity is its universality: any physical system takes part in gravitational interactions by its mere existence, i.e., by possessing a non-vanishing energy-momentum tensor. Hydrodynamics is another universal formulation in physics: any system with a large number of degrees of freedom admits a fluid description in the low-energy limit. These two universal formulations meet each other within the phase space formulation of gravity, where the induced metric—the fundamental variable of geometry—and the quasilocal energy-momentum tensor—the fundamental variable of hydrodynamics—form a canonically conjugate pair \cite{Brown:1992br}. This canonical structure implies a deep, intrinsic connection between gravitational dynamics and fluid mechanics.

The history of the connection of fluid and gravity is both long and rich. It traces back to the work of Damour, who demonstrated that the dynamics of a null horizon could be mapped to the Navier-Stokes equations \cite{Damour:1978cg}, and the ``Membrane Paradigm,'' which modeled black hole horizons as physical membranes endowed with fluid properties such as viscosity and conductivity \cite{Thorne:1986iy, Parikh:1997ma}. In the modern era, this connection was solidified within the AdS/CFT correspondence \cite{Maldacena:1997re, Aharony:1999ti, Witten:1998zw, Gubser:1998bc} as the fluid/gravity correspondence \cite{Bhattacharyya:2007vjd, Rangamani:2009xk, Hubeny:2011hd}. Here, the long-wavelength dynamics of the bulk geometry are rigorously mapped to the hydrodynamics of the dual boundary field theory.

In the standard formulation of the fluid/gravity correspondence \cite{Rangamani:2009xk, Hubeny:2011hd}, one typically considers a conformal relativistic fluid residing at the asymptotic AdS boundary. This description and, in general, the standard AdS/CFT formulation, rely heavily on the  Dirichlet boundary conditions imposed on the fields on the asymptotic AdS boundary. Dirichlet boundary conditions fix the metric at the boundary and effectively freeze the background geometry, leaving the dual stress-energy tensor as the sole dynamical variable. In this regime, the system describes a fluid flowing on a fixed, non-dynamical background, where the Navier-Stokes equation follows from the diffeomorphism invariance of the bulk theory. The fluid/gravity correspondence has been extensively studied, leading to the precise calculation of transport coefficients and the discovery of universal bounds on fluid dissipation \cite{Kovtun:2004de}.

In this work, we propose an extension of the fluid/gravity correspondence to the \textit{freelance fluid/gravity correspondence}. Building upon the ``freelance holography'' \cite{Parvizi:2025shq, Parvizi:2025wsg, Taghiloo:2025oeu, Sheikh-Jabbari:2025kjd} framework, extending gauge/gravity correspondence to the cases with arbitrary timelike boundaries (inside AdS) with arbitrary boundary conditions on it, we generalize the fluid/gravity correspondence to beyond the Dirichlet case. We specifically focus on the AdS$_3/$CFT$_2$ setup. Three-dimensional gravity offers a unique laboratory for this study: the absence of propagating bulk degrees of freedom mirrors the absence of dissipative terms in the fluid description, and the integrability of the system allows for the construction of exact solutions for flow equations. See \cite{Sheikh-Jabbari:2025kjd} for a detailed analysis of freelance holography in AdS$_3$ case. 

A central theme in our analysis is the radial evolution of the system. In the context of the gauge/gravity duality, the radial coordinate is identified with the renormalization group (RG) scale of the boundary theory \cite{Susskind:1998dq, deBoer:1999tgo, Balasubramanian:1999zv, Heemskerk:2010hk, Faulkner:2010jy}. Consequently, moving the timelike boundary from asymptotic infinity to a finite distance corresponds to a flow from the ultraviolet (UV) to the infrared (IR) in the CFT side. The freelance fluid/gravity correspondence provides the natural framework to study RG flow in the fluid description.

This perspective raises a crucial question: how does the fluid description evolve when moving to a finite cutoff \cite{Faulkner:2010jy, Bredberg:2010ky, Iqbal:2008by, Compere:2011dx, Pinzani-Fokeeva:2014cka}? While the Dirichlet condition is natural at infinity, the bulk dynamics induce a certain mixed boundary condition at any finite distance. As we will discuss, this implies that a finite-distance observer does not see a fluid on a fixed background, but rather a \textit{hydro-gravitational system}—a fluid coupled to a fluctuating, dynamical geometry. We provide a comprehensive analysis of this flow, rigorously defining the fluid description for general non-Dirichlet boundary conditions. 

\paragraph{Organization of the paper.}
Section \ref{sec:Basic-setup} reviews the necessary preliminaries of AdS$_3$ gravity required to construct the framework. In Section \ref{sec:asymptotic-Fluid/gravity--bc}, we provide a general discussion of asymptotic boundary conditions, detailing the constraints they impose on the solution phase space and analyzing the descriptions of both asymptotic timelike and null fluids. Section \ref{sec: finite-cut-fluid} constitutes the core of our analysis, where we derive the RG flow equations for the fluid variables. We solve these equations explicitly, interpolating between fluid descriptions at different energy scales. Notably, we prove the $v_g$-theorem, establishing the monotonic radial evolution of the sound group velocity. In Section \ref{sec: examples}, we apply our formalism to explicit examples, tracking the RG flow of asymptotic fluids under various specific boundary conditions. Finally, Section \ref{sec:discussion} provides a summary of our results and an outlook for future research. Appendix \ref{appen:useful-rel} contains useful relations for the RG interpolation analysis.

\section{Preliminaries}\label{sec:Basic-setup}
In this section, we present the essential background material required for the developments that follow. A more detailed discussion and analysis may be found in \cite{Sheikh-Jabbari:2025kjd}.

\subsection{\texorpdfstring{AdS$_3$}{AdS3} gravity }\label{sec:ADS-gravity}

\paragraph{Geometric setup.}
We begin by explicitly outlining the geometric framework. Our analysis takes place in the standard AdS$_3$/CFT$_2$ setting. The bulk line element is written in the Fefferman--Graham gauge \cite{Fefferman:1985abc} as
\begin{equation}\label{metric}
    \d{} s^2 = \frac{\ell^2}{r^2}\,\d{} r^2 + \frac{r^2}{\ell^{2}}\,\gamma_{ab}\,\d{} x^{a}\d{} x^{b}\,,\qquad \gamma_{ab}=\gamma_{ab}(r, x^a)\,,
\end{equation}
where $\ell$ is the AdS$_3$ radius, $r$ is the radial coordinate, and $\gamma_{ab}$ is the conformal induced metric on the timelike hypersurfaces of constant $r$. The coordinates $x^a$ label points on these hypersurfaces. 

\begin{figure}[t]
\centering
\begin{tikzpicture}[scale=1.2]

\node (v1) at (0.8,0) {};
\node (v4) at (0.8,-3) {};
\node (v5) at (-0.8,0) {};
\node (v8) at (-0.8,-3) {};

\begin{scope}
  \clip ($(v1)+(0,0)$) to[out=135,in=45, looseness=0.5] ($(v5)+(0,0)$)
        -- ($(v8)+(0,0)$) to[out=45,in=135, looseness=0.5] ($(v4)+(0,0)$)
        -- cycle;
  \fill[gray!20] (-0.8,0) -- (-0.8,-3) -- (0.8,-3) -- (0.8,0) -- cycle;
\end{scope}

\draw [darkred!60, very thick] (-0.8,0) -- (-0.8,-3);
\draw [darkred!60, very thick] (0.8,0) -- (0.8,-3);

\begin{scope}[fill opacity=0.8, very thick, darkred!60]
  \filldraw [fill=gray!30] (0.8,0) 
    to[out=135,in=45, looseness=0.5] (-0.8,0)
    to[out=315,in=225,looseness=0.5] (0.8,0);
\end{scope}
\begin{scope}[fill opacity=0.8, very thick, darkred!60]
  \filldraw [fill=gray!30] (0.8,-3) 
    to[out=135,in=45, looseness=0.5] (-0.8,-3)
    to[out=315,in=225,looseness=0.5] (0.8,-3);
\end{scope}

\draw [blue!60](0,0) ellipse (1.5 and 0.5);
\draw [blue!60] (0,-3) ellipse (1.5 and 0.5);
\draw [blue!60] (-1.5,0) -- (-1.5,-3);
\draw [blue!60] (1.5,0) -- (1.5,-3);

\fill (0,-1.5)  node [blue!50!black] {${\cal M}_r$};
\fill (0.8,-2)+(0.3,0.5) node [darkred!60] {$\Sigma_r$};
\fill (0.8,-3)+(0.7,0.8) node [right,blue!80] {$\Sigma$};
\fill (0, -3.8) node [black] { $\mathcal{M}$};
\end{tikzpicture}
\caption{ \justifying{\footnotesize{ Asymptotically AdS$_3$ spacetime and a region cutoff at radius $r$.  $\textcolor{black}{\cal{M}}$ denotes the global asymptotically AdS$_3$ spacetime and  $\textcolor{blue!80}{\Sigma}$ is its asymptotic timelike boundary.  The shaded region $\textcolor{blue!50!black}{\mathcal{M}_r}$ is the part of AdS$_3$ cutoff at radius $r$, enclosed in a timelike surface $\textcolor{darkred!60}{\Sigma_r}$.}}}
\label{fig:ADS-timelike}
\end{figure}
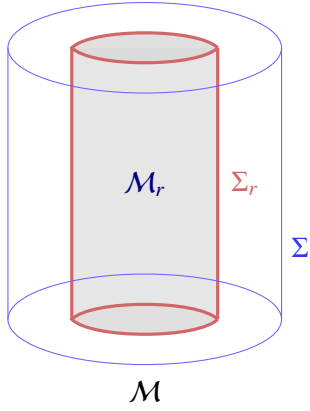

We foliate the bulk spacetime $\mathcal{M}$ by the constant-$r$ hypersurfaces $\Sigma_r$ and denote by $\mathcal{M}_r$ the interior region bounded by $\Sigma_r$; see Fig.~\ref{fig:ADS-timelike}. The asymptotic boundary of AdS is obtained in the limit $r \to \infty$, which we denote by $\Sigma \equiv \lim_{r\to\infty}\Sigma_r$. To ensure that the spacetime is asymptotically AdS, we require $\lim_{r\to\infty} \gamma_{ab} \sim \mathcal{O}(1)$.

The normal one-form to \(\Sigma_r\), together with the corresponding induced metric, are given by  
\begin{equation}
    s = s_{\mu}\,\d{} x^{\mu} = \frac{\ell}{r}\,\d{} r \,, 
    \qquad 
    h_{\mu\nu} := g_{\mu\nu} - s_{\mu}s_{\nu}\,.
\end{equation}  
The extrinsic curvature of \(\Sigma_r\) is defined as  
\begin{equation}\label{K-form}
    K_{\alpha\beta}
    = \frac{1}{2}\, h^{\mu}{}_{\alpha}\, h^{\nu}{}_{\beta}\, \mathcal{L}_{s} h_{\mu\nu}\, .
\end{equation}
Evaluating this for the metric components yields the tensor $K_{ab}$ of $\Sigma_r$ and its trace $K$
\begin{equation}
    K_{ab} = \frac{r^2}{\ell^2}\left( \ell^{-1} \gamma_{ab} + \frac{r}{2\ell}\partial_{r}\gamma_{ab} \right)\, , \qquad K= 2\ell^{-1} + \frac{r}{\ell} \frac{\partial_r \sqrt{-\gamma}}{\sqrt{-\gamma}}\, .
\end{equation}
\paragraph{Notation.}
Throughout this paper, Greek indices ($\mu, \nu, \dots$) refer to the bulk spacetime, which are raised and lowered using the full metric $g_{\mu\nu}$. Latin indices ($a, b, \dots$) denote coordinates on the hypersurface $\Sigma_r$, which are raised and lowered using the metric $\gamma_{ab}(r, x^a)$.

\paragraph{AdS$_3$ Einstein gravity action.}\label{sec:AdS3-gravity}
We focus on formulating the dynamics within the region \(\mathcal{M}_r\). The three-dimensional Einstein--Hilbert action evaluated on \(\mathcal{M}_r\) with Dirichlet boundary conditions imposed on its radial boundary \(\Sigma_r\) is,  
\begin{equation}\label{action-AdS-3}
    S[\mathcal{M}_r]
    = \frac{1}{2\kappa} \int_{\mathcal{M}_r} \!\sqrt{-g}\,\left(\mathscr{R} + \frac{2}{\ell^2}\right)
    + \frac{r^2}{\ell^2\kappa} \int_{\Sigma_r} \!\sqrt{-\gamma}\, (K - \ell^{-1})\,,
\end{equation}
where \(\kappa = 8\pi G\) and \(\mathscr{R}\) is the scalar curvature of the bulk metric \(g_{\mu\nu}\). The boundary contribution in \eqref{action-AdS-3} includes the Gibbons--Hawking--York (GHY) term \cite{PhysRevD.15.2752, PhysRevLett.28.1082} together with the standard counterterm \cite{Balasubramanian:1999re, Emparan:1999pm}. These boundary terms render the action finite and ensure a well-posed variational principle under Dirichlet boundary conditions.

Varying the action \eqref{action-AdS-3} leads to the Einstein field equations
\begin{equation}\label{eom-AdS3}
    \mathscr{R}_{\mu\nu} - \frac{1}{2}\mathscr{R}\, g_{\mu\nu} - \frac{1}{\ell^2} g_{\mu\nu} = 0\,,
\end{equation}
together with the Dirichlet symplectic potential
\begin{equation}\label{sym-pot-r}
    \Theta[\Sigma_r] 
    := \delta S_{\text{\tiny bulk}}[\mathcal{M}_r]\big|_{\text{on-shell}} 
    = -\frac{1}{2} \int_{\Sigma_r} \sqrt{-\gamma}\, \mathcal{T}^{ab}\, \delta \gamma_{ab}\,,
\end{equation}
where $\mathcal{T}_{ab}$ is the renormalized Brown--York stress tensor (rBY-EMT) \cite{Balasubramanian:1999re}, defined as
\begin{equation}\label{BY-EMT-unren}
    \mathcal{T}_{ab} = \frac{1}{\kappa}\left(K_{ab} - \frac{r^2}{\ell^2}K \gamma_{ab} + \frac{r^2}{\ell^3} \gamma_{ab}\right)\,.
\end{equation}
For practical computations, it is convenient to decompose the Einstein equations \eqref{eom-AdS3} in $1+2$ form using the Fefferman--Graham gauge \eqref{metric}:
\begin{subequations}\label{EoM-GR-mat-decompose}
\begin{align}
    &\mathcal{T} + \frac{c}{24\pi}\, \mathcal{R} + \frac{6\pi\ell^4}{c r^2}\, \Ottbar = 0\,,
    \label{EoM-ss} \\[0.2cm]
    &\nabla_b \mathcal{T}^{b}{}_{a} = 0\,,
    \label{EoM-sa} \\[0.2cm]
    &r\,\partial_r \mathcal{T}_{ab} - \frac{12\pi\ell^4}{c r^2}\left( \mathcal{T}\,\mathcal{T}_{ab} + \frac{3}{2}\, \Ottbar\, \gamma_{ab} \right) = 0\,,
    \label{EoM-ab} \\
    & r \partial_{r}\gamma_{ab} - \frac{24\pi \ell^4}{c r^2} (\mathcal{T}_{ab} + \mathcal{T} \gamma_{ab}) = 0\, ,
\end{align}
\end{subequations}
where $c := 12\pi \ell / \kappa$ is the Brown--Henneaux central charge \cite{Brown:1986nw}, $\mathcal{R}$ is the Ricci scalar of $\gamma_{ab}$ on $\Sigma_r$, $\nabla_a$ is the associated covariant derivative, and the T$\bar{\text{T}}$ operator \cite{Zamolodchikov:2004ce, Smirnov:2016lqw} is
\begin{equation}\label{ottbar}
    \Ottbar := \mathcal{T}^{ab}\mathcal{T}_{ab} - \mathcal{T}^2\,.
\end{equation}
\paragraph{Integrating radial bulk evolution.} 
In three dimensions, the radial evolution of the Einstein equations in the Fefferman--Graham gauge \eqref{metric} is \textit{integrable}. The induced metric $\gamma_{ab}(r,x)$ can be solved explicitly in terms of two integration constants, a codimension-one tensor $q_{ab}(x)$ and a boundary stress tensor $T_{ab}(x)$. In terms of the trace-reversed stress tensors,
\begin{equation}
    \tilde{\mathcal{T}}_{ab} := \mathcal{T}_{ab} - \mathcal{T}\, \gamma_{ab} \,, 
    \qquad 
    \tilde{T}_{ab} := T_{ab} - T\, q_{ab}\,,
\end{equation}
the solution for the induced metric and trace-reversed BY-EMT read \cite{McGough:2016lol, Guica:2019nzm, AliAhmad:2025kki, Sheikh-Jabbari:2025kjd}
\begin{subequations}\label{h-calT}
    \begin{align}
    & \gamma_{ab}(r,x) = 
    \Bigg( q_{ac} - \frac{6\pi \ell^4}{c r^2}\, \tilde{T}_{ac} \Bigg) 
    q^{cd} 
    \Bigg( q_{bd} - \frac{6\pi \ell^4}{c r^2}\, \tilde{T}_{bd} \Bigg)\,, \label{h-sol-summary} \\
    & \tilde{\mathcal{T}}_{ab}(r,x)  = \Big(q_{ac} - \frac{6\pi\ell^4}{c r^2}\tilde{T}_{ac} \Big) q^{cd} \tilde{T}_{db}\, , \label{calT-sol-summary}
    \end{align}
\end{subequations}
where $q^{ab}$ is the inverse of $q_{ab}$. The integration constants are subject to the constraints
\begin{equation}\label{T-constraints-summary}
    D_a T^{ab} = 0\,, \qquad 
    T := q^{ab} T_{ab} = - \frac{c}{24 \pi} R \,,
\end{equation}
where $D_a$ and $R$ are the covariant derivative and Ricci scalar associated with $q_{ab}$.  

The tensors $q_{ab}$ and $T_{ab}$ correspond to the asymptotic values of the induced metric and the renormalized Brown--York stress tensor,
\begin{equation}
    q_{ab}(x) = \lim_{r\to\infty} \gamma_{ab}(r,x)\,, 
    \qquad 
    \tilde{T}_{ab}(x) = \lim_{r\to\infty} \tilde{\mathcal{T}}_{ab}(r,x)\,.
\end{equation}
Thus, the most general solution in three dimensions is fully determined by the four independent components of $q_{ab}$ and $T_{ab}$, subject to the conservation and trace conditions \eqref{T-constraints-summary}, and also the boundary conditions that should be specified through other physical requirements.

For later convenience, we also write the inverse of \eqref{h-calT} (see Appendix~\ref{appen:useful-rel} for further details)
\begin{equation}\label{q-T-sol-summary}
        q_{ab} = {\cal H}_{ac} \gamma^{cd}{\cal H}_{db}\, , \qquad   \tilde{T}_{ab} =  {\cal H}_{ac} \gamma^{cd} \tilde{\mathcal{T}}_{db}\, ,
\end{equation}
where
\begin{equation}\label{def-calH}
    {\cal H}_{ab}:=\gamma_{ab} +\frac{6\pi\ell^4}{c r^2}\tilde{{\cal T}}_{ab}\, .
\end{equation}
\subsection{Radial flow of the solution phase space}

In this subsection, we investigate how the solution phase spaces at different radial locations are related to one another. Owing to the radial integrability of AdS$_3$ gravity, we construct an exact map between the solution spaces defined at various radii.

To do so, we begin by mapping the symplectic potentials at different radii. The critical equation is the flow of the symplectic potential
\begin{equation}
      \frac{\mathrm{d}}{\mathrm{d}r} \Theta[\Sigma_r] = \delta \int_{\Sigma_r} \mathcal{L}\Big|_{\text{on-shell}} = - \frac{12 \pi \ell^4}{c r^3} \delta\int_{\Sigma_r} \sqrt{-\gamma}\, \Ottbar\, .
\end{equation}
Using \eqref{sym-pot-r}, we find
\begin{equation}
      -\frac{1}{2}\frac{\mathrm{d}}{\mathrm{d}r} \int_{\Sigma_r}\! \sqrt{-\gamma}\, \mathcal{T}^{ab}\, \delta \gamma_{ab} = - \frac{12 \pi \ell^4}{c r^3} \delta\int_{\Sigma_r} \sqrt{-\gamma}\, \Ottbar\, .
\end{equation}
Integrating from $r=\infty$ to a finite distance $r$ we get
\begin{equation}\label{symp-pot-flow}
   \inbox{  -\frac{1}{2} \int_{\Sigma}\! \sqrt{-q}\, {T}^{ab}\, \delta q_{ab}  + \frac{1}{2} \int_{\Sigma_r}\! \sqrt{-\gamma}\, \mathcal{T}^{ab}\, \delta \gamma_{ab}  = - \frac{12 \pi \ell^4}{c}\,  \delta \int_{r}^{\infty} \frac{\d{}r}{r^3}\int_{\Sigma_r} \sqrt{-\gamma}\, \Ottbar\, .}
\end{equation}
This  equation shows that the symplectic potentials at different radial locations are related through a Legendre transformation, with $\Ottbar$ acting as the generator of this transformation. Moreover, it  shows how boundary conditions flow under RG and importantly shows that one cannot impose Dirichlet boundary conditions simultaneously at infinity and at a finite radial cutoff $r$. Once the Dirichlet condition is fixed at a given boundary, it generically induces a mixed boundary condition at any other radius \cite{Adami:2025pqr}.

Let us impose Dirichlet boundary conditions at infinity, $\delta q_{ab}=0$. Then, the radial flow equation \eqref{symp-pot-flow} enforces the following boundary condition at a finite radial cutoff $r$:
\begin{equation}\label{mixed-bc-finite}
   \inbox{ \delta\!\left( {\cal H}_{ac}\, \gamma^{cd}\, {\cal H}_{db} \right) = 0 \, .}
\end{equation}
Here, ${\cal H}_{ab}$ is defined in \eqref{def-calH}. As is evident from its definition, this condition involves both the induced metric and its conjugate momentum, and therefore corresponds to a mixed (Dirichlet–Neumann) boundary condition rather than a purely Dirichlet or Neumann one.

We are now in a position to map the solution phase space at different radial locations. This mapping can be discussed at two distinct levels: first, without requiring a well-defined variational principle—equivalently, without imposing boundary conditions—and second, at the level of a closed system with a well-defined action principle, where boundary conditions are imposed.

\begin{enumerate}
    \item[\textbf{(i)}] \textbf{Without imposing boundary conditions.}  
    The asymptotic solution phase space of the theory is characterized by the canonical pair $\{q_{ab}, T^{ab}\}$, subject to the constraints summarized in \eqref{T-constraints-summary}. These variables serve as labels of the asymptotic solution space. At a finite radial cutoff, the natural canonical data are given instead by the pair $\{\gamma_{ab}(r), \mathcal{T}^{ab}(r)\}$. A key observation is that the asymptotic and finite-radius canonical pairs are not independent. Rather, there exists an exact and invertible map between them, explicitly given by \eqref{h-sol-summary} and \eqref{calT-sol-summary}, with the inverse relations summarized in \eqref{q-T-sol-summary}.
    
    \item[\textbf{(ii)}] \textbf{With imposing boundary conditions.}  
    To obtain a well-defined solution phase space, one must impose appropriate boundary conditions that render the variational principle well posed and the system closed. A crucial point is that imposing a boundary condition at one radial location—either at infinity or at a finite cutoff—necessarily determines, or induces, a specific boundary condition at any other radius. For instance, constructing a Dirichlet solution space at infinity, where the only dynamical data are carried by $T^{ab}$, induces a mixed boundary condition at a finite radius, given explicitly in \eqref{mixed-bc-finite}. Consequently, the solution space at a finite distance is naturally described in terms of mixed variables. Physically, this shift from Dirichlet to mixed variables represents the transition from a fluid on a fixed background to a fluid coupled to induced gravity.
\end{enumerate}
\section{Freelance fluid/gravity correspondence: Various asymptotic  boundary conditions}\label{sec:asymptotic-Fluid/gravity--bc}

We initiate the construction of the freelance fluid/gravity correspondence by focusing on the asymptotic AdS boundary. While the standard formulation of the duality \cite{Bhattacharyya:2007vjd, Rangamani:2009xk, Hubeny:2011hd} relies exclusively on Dirichlet boundary conditions—describing a fluid on a fixed background—the ``freelance'' framework \cite{Parvizi:2025shq, Parvizi:2025wsg, Sheikh-Jabbari:2025kjd} allows for the imposition of arbitrary boundary conditions. Here, we analyze the structure of the asymptotic solution phase space subject to such general conditions. We systematically explore the constraints imposed by the variational principle and the gravitational equations of motion, characterizing the resulting hydrodynamic descriptions. Our analysis encompasses both timelike and null fluids, establishing the physical requirements for causality and stability at the asymptotic boundary before extending the duality into the bulk in subsequent sections.

\subsection{Asymptotic boundary conditions}

The asymptotic fluid is specified by two tensors $q_{ab}, T^{ab}$. A generic boundary condition is then specified through,
\begin{equation}\label{asymp-bc}
\delta \bcB_{ab}=0 \, , \qquad 
\bcB_{ab} = \bcN\, q_{ab} + \bcD \, T_{ab}\, ,
\end{equation}
where the functions $\bcD$ and $\bcN$ are allowed to depend on the boundary metric $q_{ab}$ and on the combination $\sqrt{-q}\,T^{ab}$, namely $\bcD=\bcD(q_{ab},\sqrt{-q}\,T^{ab})$ and
$\bcN=\bcN(q_{ab},\sqrt{-q}\,T^{ab})$. It is seen that the $\bcD=0$ case yields Dirichlet boundary conditions and $\bcN=0$ Neumann boundary conditions. 

A given boundary condition $\delta \bcB_{ab}=0$ can be made compatible with  the variational principle if there exists a tensor $\bcC^{ab}=\bcC^{ab}(q_{ab},\sqrt{-q}\, T^{ab})$ such that, 
\begin{equation}\label{well-action-prin}
\int_{\Sigma} \delta(\sqrt{-q}\, T^{ab}) \wedge\,\delta q_{ab}
= \int_{\Sigma} \delta \bcC^{ab}\,\wedge \delta \bcB_{ab} + \text{corner terms} \, .
\end{equation}
The above implies that there should be a canonical (Legendre) transformation from $(q_{ab}, \sqrt{-q}T^{ab})$ to $(\bcB_{ab}, \bcC^{ab})$. Eq.\eqref{well-action-prin} may also be written as
\begin{equation}\label{well-action-prin-W-Y}
 \bcC^{ab}\,\delta \bcB_{ab} =  \ \sqrt{-q}\, T^{ab} \,\delta q_{ab}+  \delta W + \partial_a Y^a \, ,
\end{equation}
for a vector $Y^a$ which is a one-form over the solution space and a function $W=W(q_{ab},\sqrt{-q}\,T^{ab})$. The latter is the Legendre-generating-functional implementing the transformation from Dirichlet boundary conditions to the more general condition $\delta \bcB_{ab}=0$. 

The integrability of \eqref{well-action-prin-W-Y} then implies the following conditions:
\begin{equation}
\begin{split}
\sqrt{-q}\,T^{ab}
&= \bcC^{cd}\,\frac{\delta \bcB_{cd}}{\delta q_{ab}}
   + \frac{\delta W}{\delta q_{ab}} + \frac{\delta (\partial_c Y^c)}{\delta q_{ab}}\, , \\
0
&= \bcC^{cd}\,\frac{\delta \bcB_{cd}}{\delta(\sqrt{-q}\,T^{ab})}
   + \frac{\delta W}{\delta(\sqrt{-q}\,T^{ab})} + \frac{\delta ( \partial_c  Y^c)}{\delta(\sqrt{-q}\,T^{ab})} \, .
\end{split}
\end{equation}
From these equations, one may solve for $\bcC^{ab}, W, Y^a$ for a prescribed choice of $\bcB_{ab}$. If the above admit a solution, we say that the corresponding boundary condition \eqref{asymp-bc} is compatible with a well-defined variational (action) principle.
\subsection{Asymptotic constraints}
Besides the boundary conditions, the asymptotic boundary data are required to satisfy the Hamiltonian and momentum constraints summarized in \eqref{T-constraints-summary},
\begin{subequations}
\begin{align}
& T = - \frac{c}{24\pi}\, R \,, \qquad 
\delta T = - \frac{c}{24\pi}\, \delta R \, , \label{var-Hamil-const} \\
& D_a T^{ab} = 0 \,, \qquad 
\delta \!\left( D_a T^{ab} \right) = 0 \, . \label{var-mom-const}
\end{align}
\end{subequations}
Consequently, the boundary data $\{q_{ab},T^{ab}\}$ are subject to three boundary conditions \eqref{asymp-bc}, one scalar (Hamiltonian) constraint \eqref{var-Hamil-const}, and two differential (momentum) constraints \eqref{var-mom-const}. Altogether, these relations fix the boundary data $\{q_{ab}, T^{ab}\}$ up to two independent codimension-two corner degrees of freedom; that is, two functions of one variable in three dimensions \cite{Sheikh-Jabbari:2025kjd}.
\subsection{Asymptotic fluid}\label{subsec: asymp-fluid}
We begin by noting that the two-dimensional renormalized holographic energy--momentum tensor $T^{ab}$, defined on the asymptotic boundary $\Sigma$, can be written in the form of a perfect fluid,
\begin{equation}\label{asymp-perfect-fluid}
T_{ab} = (\rho + p)\, u_a u_b + p\, q_{ab}\, , 
\qquad q_{ab} u^a u^b = -1 \, ,
\end{equation}
where $\rho$, $p$, and $u^a$ denote the energy density, pressure, and fluid velocity, respectively; $u^a$ is a unit-norm future-oriented  timelike vector field. 
In terms of these fluid variables, the trace of the stress tensor and the $\mathrm{T}\bar{\mathrm{T}}$ operator take the form
\begin{equation}
T = p - \rho \, , \qquad 
\ttbarb = 2 \rho p \, .
\end{equation}
The solution space is specified by the (boundary variables) $\rho$, $p$, $u_a$, and $q_{ab}$. However, these are not independent. Conservation of the energy--momentum tensor,
$D^a T_{ab} = 0$, may be written as, 
\begin{subequations}\label{Euler-eq}
\begin{align}
(\rho + p)\, u^a  D_a u^b &= -(u^a u^b+q^{ab}) D_a p\, , \\
(\rho + p)\,  D_au^a & =-u^a D_a \rho \,.
\end{align}
\end{subequations}
In addition, the energy density and pressure are related by the holographic \textit{asymptotic fluid equation of state},
\begin{equation}\label{asymp-EoS}
p = \rho - \frac{c}{24\pi}\, R \, , 
\qquad R = R(q_{ab}) \, .
\end{equation}
The null energy condition (NEC) then implies that, 
\begin{equation}\label{NEC-asympt}
    \rho+p \geq 0, \qquad \Longrightarrow \qquad \rho\geq \frac{c}{48\pi} R\,.
\end{equation}
In addition, all physical configurations should satisfy the desired/prescribed boundary conditions $\delta\bcB_{ab}=0$,
\begin{equation}
    \delta \!\left[ (\bcN + \bcD\, p)\, q_{ab}
+ \bcD (\rho + p) u_a u_b \right] = 0 \, ,
\end{equation}
as well as 
\begin{subequations}\label{asymp-ensemble}
\begin{align}
& \delta p = \delta \rho - \frac{c}{24\pi}\, \delta R \, , \\
& \delta \!\left[(\rho + p) u^a D_a u_b\right]
= - \delta(D_b  p )
  - \delta \bigg\{ D_a \!\left[(\rho + p) u^a\right] u_b \bigg\} \, , 
\end{align}
\end{subequations}
which are variations of \eqref{Euler-eq} and \eqref{asymp-EoS}. These equations define the solution phase space associated with the boundary condition $\delta\bcB_{ab}=0$, which we refer to as ${\cal P}_{\bcB}$.

\paragraph{Fluid in null frame and null fluids.}
In \eqref{asymp-perfect-fluid} we considered a fluid with timelike velocity vector field $u^a$. Depending on the boundary conditions, we may be dealing with a fluid with a null velocity field, a Carrollian fluid \cite{Ciambelli:2018wre, deBoer:2017ing}. The energy momentum tensor of such a $2d$ fluid may be obtained as a specific limit of the one described in \eqref{asymp-perfect-fluid}. To facilitate taking the limit, we  rewrite $u^a$ in terms of future-oriented null vectors $l^a, n^a$:
\begin{equation}
\begin{split}
    u^a= \beta l^a+ \beta^{-1} n^a\, , &\qquad {q_{ab} = -2 (l_{a} n_{b} + l_{b} n_{a})} \, ,\\
    l^2=0=n^2\, , &\qquad \qquad l\cdot n=-1/2\, ,
\end{split}
\end{equation}
yielding
\begin{equation}\label{asymp-perfect-null-fluid}
T_{ab} 
=(\rho + p)\, (\beta^2 l_a l_b+ \beta^{-2} n_an_b)+ \frac{c}{24\pi}  (l_a n_b+  n_a l_b) R\, , 
\end{equation}
where we used \eqref{asymp-EoS} and $\beta\geq 0$ may be viewed as the boost factor. 

In the $\beta\to \infty$, we find the null fluid. In this limit, which in $2d$ coincides with $\beta\to 0$ limit, 
\begin{equation}\label{null-fluid-T}
    T_{ab} =  {\cal E}\, l_a l_b+\frac{c}{24\pi}  (l_a n_b+  n_a l_b) R\, , \qquad {\cal E}:= (\rho + p)\beta^2= \text{fixed}\, , \qquad \beta\to\infty\, .
\end{equation}
For this case, \eqref{Euler-eq} take the form
\begin{equation}\label{Euler-eq-null}
     D_a ({\cal E}\, l^a) +2 {\cal E}\, l^a l^b D_a n_b { +} \frac{c}{{24}\pi} n^a D_a R=0\, , \qquad l^a D_a R=0 \,.
\end{equation}
The null energy condition implies ${\cal E}\geq 0$ and we have, 
\begin{equation}
    T=-\frac{c}{24\pi} R,\qquad \ttbarb= -\frac12T^2=-\frac12 \left(\frac{c}{24\pi}\right)^2 R^2\, .
\end{equation}
Therefore, for flat boundary $R=0$ case, $ T_{ab} =  {\cal E}\, l_a l_b$ and both $T, \ttbarb$ vanish. 

\paragraph{Discriminant of timelike and null fluids.}
Consider $\Delta (r):$
\begin{equation}\label{discriminant-def}
    \Delta^2(r) := 2 {\cal T}^{ab}{\cal T}_{ab} - {\cal T}^2 = 2\Ottbar + {\cal T}^2\, . 
\end{equation}
For the asymptotic fluid with timelike  \eqref{asymp-perfect-fluid} and null \eqref{null-fluid-T} decompositions, we respectively have
\begin{equation}\label{discriminant-fluid-asym}
    \begin{split}
         \text{timelike:} & \qquad  \Delta^2 = (\rho + p)^2 > 0\, , \\ 
         \text{null:} & \qquad \Delta^2 = 0\, ,
    \end{split}
\end{equation}
where $\Delta=\Delta(r=\infty)$. So, $\Delta$ determines whether we have a timelike or null fluid. {The vanishing of $\Delta$ in the null case is compatible with Eq.~\eqref{null-fluid-T}, which requires taking the limit $(\rho + p) \to 0$.}

\subsection{Causal propagation of sound waves}

The fluid description is physically valid when certain stability conditions are fulfilled. Here, we discuss the one necessitated by the subluminality of the propagation of waves in the dual CFT. The speed of sound in the fluid, the group velocity of linear fluid perturbations, is given by
\begin{equation}
{v_g^2 := 
\left.\frac{\delta p}{\delta \rho}\right|_{{\cal P}_{\bcB}}\, }
\end{equation}
where the subscript ${\cal P}_{\bcB}$ indicates that the variation is taken on the solution space ${\cal P}_{\bcB}$. Stability of the fluid description, and hence physicality of a given boundary condition,  requires $v_g^2\leq 1$. 

As we will discuss below, whether the subluminality condition $v_g^2\leq 1$ is fulfilled, in general depends on the boundary conditions as well as on whether the boundary is curved. If the boundary conditions allow for  the energy density fluctuations, $\delta\rho \neq 0$, from \eqref{asymp-EoS}, one learns, 
\begin{equation}
v_g^2 = \left. 1 -\frac{c}{24\pi} \frac{\delta R}{\delta \rho}\right|_{{\cal P}_{\bcB}}\,.
\end{equation}
Thus, if  $\frac{\delta R}{\delta \rho}\geq 0$ over the solution space, $v_g^2\leq 1$, and otherwise we have superluminal instability. However, if the boundary conditions (see analysis of conformal and Neumann boundary conditions in sections \ref{sec:CBC} and \ref{sec:NBC}) do not allow for fluctuations in $\rho, p$ and $\delta\rho, \delta p$  vanish on ${{\cal P}_{\bcB}}$, the group velocity is zero.

\subsection{Flat asymptotic boundary}
For the special case of a flat asymptotic boundary, $R=0$, the equation of state reduces to $p=\rho$, and consequently 
\begin{subequations}\label{Euler-eq-R0}
\begin{align}
T_{ab} &= (2u_a u_b+ q_{ab}) \rho\, , \\ 
(q^{ab}+ u^a u^b)\frac{D_b \rho}{2\rho} &= -u^b D_b u^a\, , \qquad D_a(\sqrt{\rho}\, u^a) = 0 \, .
\end{align}
\end{subequations}
For this case, NEC implies $\rho\geq 0$. If the adopted boundary conditions allow for $\delta\rho\neq 0$, the group velocity satisfies $v_g^2=1$, and if the adopted boundary conditions require $\delta\rho=0$,  $v_g^2=0$. That is, for the generic case, the fluid is a relativistic one residing on the AdS$_3$ boundary, which is locally a flat space. For the  $\delta\rho=0$, we are dealing with a Carrollian fluid in which there is no propagation of sound waves. 

Having established the hydrodynamic description at the asymptotic boundary, we now proceed to investigate the fluid dynamics at a finite radial cutoff in the next section.
\section{Freelance fluid/gravity correspondence: Fluid at finite cutoff }\label{sec: finite-cut-fluid}
We now examine the fluid/gravity correspondence from the perspective of radial evolution. Building on the asymptotic analysis of the previous section, we track how the fluid variables and their defining boundary conditions evolve as the boundary  is moved from infinity into the bulk. Our approach is constructive: we start with the asymptotic fluid configurations subject to general boundary conditions and utilize the bulk dynamics to transport this description to a finite cutoff surface $\Sigma_r$.

This radial flow reveals a fundamental shift in the physical nature of the system. At the asymptotic boundary (UV), a Dirichlet boundary condition ($\delta q_{ab}=0$) is natural, describing a fluid flowing on a fixed, non-dynamical background geometry. As we flow toward the finite distance (IR), the induced boundary condition generically becomes mixed. Physically, this implies that for a finite-distance observer, the background metric $\gamma_{ab}$ is no longer rigid; it fluctuates in response to the fluid's dynamics ($\delta \gamma_{ab} \neq 0$). Thus, the effective theory on $\Sigma_r$ transforms from pure hydrodynamics into a \textit{hydro-gravitational system}: a fluid coupled to 2D dynamical gravity. In the following, we quantify this transition by deriving the explicit RG flow equations for the fluid variables.

Analogous to the fluid description at infinity, the renormalized Brown--York energy--momentum tensor at a finite radial position $r$, denoted by $\mathcal{T}_{ab}$, also admits a perfect-fluid decomposition. 
The energy--momentum tensor takes the form
\begin{equation}
\mathcal{T}_{ab}(r)
= \big[\rho(r)+p(r)\big]\, u_a(r) u_b(r) + p(r)\, \gamma_{ab}(r)\, ,
\end{equation}
with the normalization condition
\begin{equation}
\gamma_{ab}(r)\, u^a(r) u^b(r) = -1 \, .
\end{equation}
The corresponding trace and $\Ottbar$ operator are
\begin{equation}
\mathcal{T}(r) = p(r) - \rho(r)\, , \qquad
\Ottbar(r) = 2\, \rho(r)\, p(r)\, .
\end{equation}
The variables $\rho(r), p(r), u^a(r)$ and $\gamma_{ab}(r)$ after imposing the equations of motion \eqref{EoM-ss}, define the solution space. We note that besides \eqref{EoM-ss}, the variation of these equations should also be satisfied on the solution space.  

Using the equations in appendix \ref{appen:useful-rel} and the analysis in \cite{Sheikh-Jabbari:2025kjd}, one may solve \eqref{EoM-ss} and obtain the \textit{equation of state at finite $r$}:
\begin{equation}\label{EoS-r}
\inbox{
p(r)
= \frac{\rho(r) - \frac{c r^2}{24\pi \ell^2}\, \mathcal{R}}{1 + \frac{12\pi \ell^{4}}{c r^2}\, \rho(r)}
\, , \qquad \mathcal{R}={\cal R}(\gamma_{ab})\ .}
\end{equation}
The above at $r\to\infty$ recovers \eqref{asymp-EoS}. It shows the RG evolution of the equation of state of the fluid, how the equation of state changes as we move in $r$.
From this equation of state, one can compute the local group velocity of sound,
\begin{equation}\label{v-g2--r}
v_g^{2}(r)
:= \left.\frac{\delta p(r)}{\delta \rho(r)}\right|_{{\cal P}_{\bcB}(r)} \, ,
\end{equation}
{where ${\cal P}_{\bcB}(r)$ denotes the solution space obtained by flowing the asymptotic solution space from the AdS boundary to the radial position $r$. Consequently, the computation of the group velocity at finite cutoff requires imposing these deformed constraints.}

\paragraph{Special case $\mathcal{R}=0$.}
If the asymptotic boundary is flat, then in three dimensions---working in Fefferman--Graham gauge---the induced metric on any constant-radius slice is also flat, cf. \eqref{r-dep-ttbar-3}. In this case, the group velocity takes the simple form
\begin{equation}\label{vg-R=0}
v_g(r) = \frac{1}{\Bigl|1 + \frac{12\pi \ell^{4}}{c\, r^2}\, \rho(r)\Bigr|} \, v_g\,.
\end{equation}
The above indicates that regardless of the boundary conditions, NEC implies $\rho (r)\geq 0$,  hence NEC yields $v_g(r) \leq 1$ and the sound waves remain subluminal at all $r$.\footnote{As pointed out earlier, boundary conditions may require, $\delta\rho=0$ (cf. sections \ref{sec:CBC} and \ref{sec:NBC}). For these cases, while still subluminal, $v_g(r)=v_g=0$.} As we will prove below, this result generalizes to cases with $R\neq 0$.

\subsection{RG flow in timelike fluids}\label{subsec: interpolation}
The RG flow of the hydrodynamic description corresponds, from the bulk perspective, to the radial evolution of the fluid variables from the asymptotic boundary to a finite cutoff. The resulting flow equations take a remarkably simple form:
\begin{equation}
\inbox{
    \begin{split}
         r \partial_{r} \rho(r) & = -\frac{12\pi\ell^4}{c r^2}\, \rho(r)^2\, , \\
         r \partial_{r} p(r) & = +\frac{12\pi\ell^4}{c r^2}\, p(r)^2\, , \\
         r \partial_{r} u_{a}(r) & = - \frac{12\pi\ell^4}{c r^2}\, p(r)\, u_{a}(r)\, .
    \end{split}}
\end{equation}
This structure is consistent with the expectation that the flow is driven by the $\Ottbar$ deformation. Integrating this system, together with the metric flow equations (see \cite{Sheikh-Jabbari:2025kjd} and Appendix \ref{appen:useful-rel}), yields the exact interpolation for the induced metric:
\begin{equation}
    \begin{split}
    \sqrt{-\gamma} & = \sqrt{-q}\, \big|A(r)\, B(r)\big|\, ,\\
    \gamma_{ab} & = A(r)^2 q_{ab}+ (A(r)^2-B(r)^2)\, u_a u_b \, , \\
    \gamma^{ab} & = A(r)^{-2} q^{ab}+ (A(r)^{-2}-B(r)^{-2})\, u^a u^b \, ,
    \end{split}
\end{equation}
and the corresponding fluid variables:\footnote{{Throughout the rest of this paper, we adopt the convention that $X(r)$ denotes a quantity defined on the finite cutoff surface $\Sigma_r$, while $X$ without an argument refers to its asymptotic value on the AdS boundary $\Sigma$, i.e., $X \equiv \lim_{r\to\infty} X(r)$.}}
\begin{equation}\label{interpolate-fluid}
\begin{split}
& p(r) = \frac{ p}{B(r)}\, , \qquad \rho(r) = \frac{\rho}{A(r)}\, , \\
& u_a(r)=  B(r) \, u_a\,, \qquad u^a(r)= B(r)^{-1} \ u^a\, ,
\end{split}
\end{equation}
where $A(r)$ and $B(r)$ are defined as follows
\begin{equation}\label{AB-def}
    A(r) :=1-\frac{6\pi \ell^4}{c r^2}\rho\, , \qquad B(r) := 1+\frac{6\pi \ell^4}{c r^2} p\, .
\end{equation}
We note that $A(r=\infty)=1=B(r=\infty)$ and thus, $\rho(r=\infty)=\rho$, $p(r=\infty)=p$, and $u_{a}(r=\infty)=u_{a}$.

Some comments are in order:
\begin{enumerate}
\item In general $\rho, p$, while $r$-independent, have $x^a$-dependence. This $x^a$-dependence depends on the boundary conditions. 
\item One may invert \eqref{interpolate-fluid} and write $\rho, p$ in terms of $\rho(r), p(r)$:
\begin{equation}
    \rho = \frac{\rho(r)}{1+\frac{6\pi \ell^4}{c r^2}\rho(r)}, \qquad  p = \frac{p(r)}{1+\frac{6\pi \ell^4}{c r^2}p(r)}.
\end{equation}
\item Eigenvalues of the metric $\gamma_{ab}$ are proportional to $A^2, B^2$. So, the $\gamma_{ab}$ metric becomes singular when $A(r)$ or $B(r)$ vanish. However, one should note that since in general $\rho, p$ have $x^a$ dependence, $A(r)=0$ or $B(r)=0$ does not necessarily happen on constant $r$ surfaces (on which $\gamma_{ab}$ is the induced metric). 
    \item The intriguing and important feature one can see from the above is that $p(r)$ is only a function of $p$ (pressure at infinity) and $\rho (r)$ is only a function of $\rho$ (energy density at infinity). This feature yields the $v_g$-theorem discussed below in section \ref{sec:vg-theorem}.  
\item From the above expressions, one can simply find
\begin{equation}\label{NEC-r-asympt}
    \sqrt{-\gamma}\ \big(\rho(r)+ p(r)\big) = \sqrt{-q}\ {\cal S} \ (\rho+p)\, , \qquad {\cal S}:=\frac{|A(r)B(r)|}{A(r)B(r)}\, . 
\end{equation}
As we see, $\sqrt{-\gamma} \left(\rho(r)+ p(r)\right)$ is essentially an $r$ independent combination. 
\item\label{item-6} The sign ${\cal S}$ in \eqref{NEC-r-asympt} is $+1$ at large $r$ and it may change sign as we go to lower $r$, where $A(r)B(r)$ may become negative. That is, if the null energy condition is satisfied by the asymptotic fluid, then it will be satisfied at any arbitrary distance as long as $A(r)B(r)$ remains positive. NEC at large $r$ implies ${\rho\geq \frac{c}{48\pi} R}$ and NEC is respected at $r$ if this condition and $A(r)B(r)>0$ are fulfilled. This happens when ${\rho\geq \frac{c}{48\pi} R}$ and
\begin{equation}\label{r-range}
    r^2\geq \frac{6\pi \ell^4}{c} \rho\, , \qquad \text{{or}} \qquad  r^2\leq \frac{6\pi \ell^4}{c} \Big(\frac{c }{24\pi}R-\rho \Big)\, .
\end{equation}

\item One can show that the following two combinations are $r$-independent:
\begin{equation}
    p(r) u_a(r) = p u_a\, , \qquad \rho(r)^2 \big(\gamma_{ab}+ u_a(r) u_b(r) \big) = \rho^2 (q_{ab}+ u_a u_b)\, .
\end{equation}
\item If $p>0$, then $p(r)>0$, and in addition $p(r)<p$. If $\rho\geq 0$, $\rho(r)$ may be positive or negative:
\begin{equation}
    \rho(r)=\frac{c}{6\pi\ell^2} \ \frac{r^2 r_c^2}{r^2-r_c^2}\,,\qquad r^2_c:=\frac{6\pi \ell^4}{c}\rho\,.
\end{equation}
Hence, for $r\geq r_c$, $\rho(r)>0$ and for $r<r_c$, $\rho(r)<0$. $\rho(r_c)$ blows up  at $r_c$. 

\end{enumerate}

\subsection{$v_g$-theorem} \label{sec:vg-theorem}

Using \eqref{interpolate-fluid}, one can compute the  group velocity of sound waves at a given radius $v_g(r)$ in terms of the asymptotic group velocity $v_g$:
\begin{equation}\label{vg-r-generic}
    v_g(r)=  \left|\frac{1-\frac{6\pi \ell^4}{c r^2}\rho}{1+\frac{6\pi \ell^4}{c r^2}p}\right| \ 
    v_g\, , \qquad v_g^2=1-\frac{\delta R}{\delta \rho}\, ,
\end{equation}
and thus $v_g=1$ for the $R=0$ case, unless we are dealing with boundary conditions for which $p, \rho$ are constant over the solution space. For these cases $v_g=0$; (e.g., see sections \ref{sec:CBC} and \ref{sec:NBC}). In general, however, $v_g$ differs from 1 by the amount that depends on $R$ and the boundary conditions. To avoid superluminal propagation we need to require $\frac{\delta R}{\delta \rho}\geq 0$.

From the above, one obtains
\begin{equation}
   \inbox{ \frac{\d{}v_g(r)}{\d{} r}=   
    \frac{{12}\pi \ell^4}{c r^3}\ \big(\rho(r)+ p(r) \big)\ v_g(r)\, .}
\end{equation}
Thus, the null energy condition $\rho (r)+p(r) \geq 0$ implies $\frac{\d{}v_g(r)}{\d{} r} \geq 0$. However, as discussed above for the range of $r$ where $A(r)B(r)\geq 0$, NEC at generic $r$ is implied by the NEC at asymptotic large $r$ boundary, cf. \eqref{NEC-r-asympt}. Therefore, we arrive at the statement of the $v_g$ theorem:
\begin{center}\textit{NEC implies that $v_g(r)$ is a monotonically decreasing function as we move away from the boundary and if $v_g\leq 1$ at the asymptotic boundary, $v_g(r)\leq 1$ at any finite $r$ in the range in \eqref{r-range}.}
\end{center} 
We note that this result is \textit{independent} of the boundary conditions on the fields. For the special (but not exceptional) cases where by the choice of boundary conditions, $\delta p, \delta \rho$ vanish (e.g., see sections \ref{sec:CBC} and \ref{sec:NBC}), $v_g(r)=v_g=0$ and hence the above statement is still true. 

\paragraph{Phase velocity.} 
The phase velocity for sound waves in a fluid with density $\rho(r)$ and pressure $p(r)$, is given as
\begin{equation}
    v^2_p(r):=\frac{p(r)}{\rho(r)} =  \frac{1-\frac{6\pi \ell^4}{r^2c}\rho}{1+\frac{6\pi \ell^4}{r^2c} p}\ v_p^2, \qquad v_p^2:= \frac{p}{\rho} = 1-\frac{R}{\rho}.
\end{equation}
The phase velocity is well-defined only if $\rho(r), p(r)$ have the same sign for any $r$, including the asymptotic large $r$ case. This requirement is fulfilled where $AB>0$, cf. item \ref{item-6} above. For $R=0$ cases, $v_p=1$. Note that, unlike $v_g(r)$, $v_p$ need not necessarily be subluminal and in general,
\begin{equation}
     \frac{v^2_p(r)}{v_g(r)}= \frac{v^2_p}{v_g}.
\end{equation}
Thus, similarly to $v_g(r)$, $v_p(r)$ is a decreasing function as we move to lower $r$.

\subsection{RG flow in null fluid description}
Repeating the analysis in section \ref{subsec: asymp-fluid}, one observes that  a null fluid is mapped to another null fluid under RG. The RG flow equations for a null fluid are as follows
\begin{equation}
    \inbox{\begin{split}
        & r \partial_{r} \mathcal{E}(r) = \frac{24\pi \ell^2}{c\, r^2} \mathcal{T}(r)\, \mathcal{E}(r)\, , \\
        & r \partial_{r} \mathcal{T}(r) = \frac{6\pi \ell^4}{c\, r^2}   \mathcal{T}(r)^2\, , \\
        & r \partial_{r} l_{a}(r) = - \frac{12 \pi \ell^6}{c\, r^2} \mathcal{T}(r)\, l_{a}(r)\, , \\
        & r \partial_{r} n_{a}(r) = - \frac{6\pi \ell^6}{c\, r^2} \mathcal{E}(r)\, l_{a}(r)\, .
    \end{split}}
\end{equation}
The above may be solved  as 
\begin{equation}
    \mathcal{E}(r) = \frac{\mathcal{E}}{\mathcal{A}^4}\, , \qquad \mathcal{T}(r) = \frac{T}{\mathcal{A}}\, ,  \qquad l_{a}(r) = \mathcal{A}^2\, l_{a}\, , \qquad n_{a}(r)  = n_a + \frac{\mathcal{B}}{\mathcal{A}} l_{a}\, ,
\end{equation}
where 
\begin{equation}
    \mathcal{A} = 1 + \frac{3\pi \ell^4}{c\, r^2} T = 1 - \frac{\ell^2}{8r^2}R\, , \qquad \mathcal{B} = \frac{3\pi \ell^4}{c\, r^2} \mathcal{E}\, .
\end{equation}
We note that $\mathcal{A}(r=\infty)=1$ and $\mathcal{B}(r=\infty)=0$. Note also that discriminant $\Delta$ \eqref{discriminant-def} has the following property
\begin{equation}
   \sqrt{-\gamma}\, \Delta(r) = \sqrt{-q}\, \Delta\, .
\end{equation}
This equation guarantees that the timelike and null nature of the fluid is preserved on the RG flow (radial evolution).

\section{Examples with different boundary conditions}\label{sec: examples}
So far we developed a general freelance fluid/gravity framework without imposing specific boundary conditions. In this section,  we work through  fluid description for solution phase spaces associated with four different asymptotic boundary conditions.  
\subsection{Asymptotic Dirichlet fluid}
We start with  the fluid description with Dirichlet boundary conditions at infinity, i.e. the Ba\~nados fluid \cite{Banados:1998gg, Sheikh-Jabbari:2025kjd}
\begin{equation}
    \delta q_{ab} = 0\, .
\end{equation}
In this case, we have a standard fluid description at infinity, a relativistic hydrodynamics on a fixed non-dynamical background. The solution for constraints \eqref{T-constraints-summary} with the Dirichlet boundary condition is given by
\begin{equation}
    \d{}s^2 = 2 e^{\phi} \d{}x^+ \d{}x^{-}\, ,
\end{equation}
where $\phi=\phi(x^+,x^-)$. For a given $\phi$, this metric satisfies the  Dirichlet boundary condition at infinity. The constraint equations may be solved as,
\begin{equation}\label{T-V-Dir-phi}
    \begin{split}
        & T_{\pm \pm} =L_{\pm}(x^{\pm}) +\frac{c}{48\pi} \left[ (\partial_{\pm}\phi)^2 - 2 \partial_{\pm}^2 \phi \right]\,, \\ 
        &T_{+-} = \frac{c}{24 \pi} \partial_{+}\partial_{-} \phi\, .
    \end{split}
\end{equation}
We note that, compared to the usual Ba\~nados geometries \cite{Banados:1998gg}, we allow the boundary to be non-flat. 
It is clear that the variational principle (i.e., the symplectic potential) cannot vanish unless $\delta\phi = 0$, meaning that $\phi$ is fixed and $\delta\phi=0$ over the solution space. 

For this case 
\begin{equation}\label{scalars-asymp-Dir-fluid}
\begin{split}
R=-\frac{24\pi}{c} &T= -2 e^{-\phi} \partial_{+}\partial_{-} \phi\, , \qquad \sqrt{-q}= e^{\phi}\, , \\ \ttbarb=2 &e^{-2\phi}\left(T_{++}T_{--}- (T_{+-})^2\right)\, .
\end{split}
\end{equation}
Then, we can determine the fluid variables for the perfect fluid at infinity explicitly as follows
\begin{equation}
    \begin{split}
        & \rho= e^{-\phi}\left( - T_{+-}+\sqrt{T_{++}\ T_{--}}\right)\, ,\\
        & p= e^{-\phi}\left( + T_{+-}+\sqrt{T_{++}\ T_{--}}\right)\, , 
    \end{split}
\end{equation}
and
\begin{equation}\label{fluid-velocity-D}
    u_{a} \d{}x^a = \frac{e^{\phi/2}}{\sqrt{2}}(\tilde{u}_+ \d{}x^+ +\tilde{u}_- \d{}x^-)\, , \qquad \tilde{u}_\pm :=\pm \left(\frac{T_{++}}{T_{--}}\right)^{\pm\frac14}\,,
\end{equation}
where $T_{\pm\pm}, T_{+-}$ are given in \eqref{T-V-Dir-phi}. 

Having the explicit fluid description in hand, we now compute the group and phase velocities
\begin{equation}
\begin{split}
    v_{g}^2 &= \frac{\delta p}{\delta \rho}\Big|_{\phi} = 1\, , \\
    v_{p}^2 &= \frac{p}{\rho} = \frac{+ T_{+-}+\sqrt{T_{++}\ T_{--}}}{- T_{+-}+\sqrt{T_{++}\ T_{--}}}\, .
\end{split}\end{equation}
From \(v_g = 1\), we conclude that the fluid at the AdS boundary is relativistic. We now turn to the phase velocity. For a flat AdS boundary, where \(\phi = 0\), we find \(v_p = 1\). In contrast, for a non-flat boundary, one generally has \(v_p \neq 1\).

From the expression above, we see that when \(T_{+-} \sim \partial_{+}\partial_{-}\phi < 0\)—which, via \eqref{scalars-asymp-Dir-fluid}, corresponds to positive curvature \(R > 0\)—the phase propagation is subluminal, \(v_p < 1\). Conversely, when \(T_{+-} > 0\), corresponding to negative curvature \(R < 0\), the phase velocity becomes superluminal, \(v_p > 1\). The superluminal phase propagation is a direct consequence of the violation of DEC, as $p>\rho$.

\paragraph{Finite cutoff fluid.}
To study the induced fluid description at a finite distance we start with the induced metric at an arbitrary radius $r$, 
\begin{subequations}
\begin{align}
\gamma_{\pm \pm} & = - \frac{12\pi\ell^4}{c r^2}  \Big( 1 + \frac{6\pi\ell^4}{c r^2} e^{-\phi} T_{+-} \Big) \ T_{\pm\pm}\, ,\\ 
\gamma_{+-} & =  e^{\phi} \bigg[\Big( 1 + \frac{6\pi\ell^4}{c r^2} e^{-\phi} T_{+-}\Big)^2 + { e^{-2\phi} } \left(\frac{6\pi\ell^4}{c r^2}\right)^2  T_{++} T_{--}\bigg]\, .
\end{align}
\end{subequations}
The induced ensemble at a finite distance is given as follows \cite{Sheikh-Jabbari:2025kjd}
\begin{equation}\label{D-into-DN}
    \delta(\sqrt{-\gamma}\, {\cal T}^{ab})= -\frac{c r^2}{24\pi \ell^4 } \sqrt{-\gamma}\, {\cal G}^{abcd} \delta \gamma_{cd}\, , \qquad {\cal G}^{abcd}:= \gamma^{ac} \gamma^{bd} + \gamma^{ad} \gamma^{bc} - \gamma^{ab} \gamma^{cd}\, ,
\end{equation}
where ${\cal G}^{abcd}$ is the Wheeler-De Witt metric \cite{PhysRev.160.1113}. The above shows how the Dirichlet boundary condition at large $r$ evolves as we move to generic $r$, where we get a mixture of Dirichlet and Neumann boundary conditions.

\paragraph{Group velocity.} The group velocity analysis of section \ref{sec:vg-theorem} was generic and independent of the curvature $R$ or the boundary conditions. So, those results apply to this case too, in particular, $v_g(r)$ is given by \eqref{v-g2--r}, which may also be written as
\begin{equation}
   \inbox{ v_g^{2}(r)= \frac{\delta p(r)}{\delta \rho(r)}\Big|_{R}= \left(\frac{1-\dfrac{6\pi \ell^{4}}{c r^{2}}\,p(r)}{1+\dfrac{6\pi \ell^{4}}{c r^{2}}\,\rho(r)}\right)^{2}\, }
\end{equation}
which is subluminal for any $r$, regardless of $\phi$. 

\subsection{Asymptotic conformal fluid}\label{sec:CBC}
The asymptotic conformal boundary conditions is defined through,
\begin{equation}
    \delta \Big( \frac{q_{ab}}{\sqrt{-q}} \Big)=0 \, , \qquad \delta T = 0 \, .
\end{equation}
From the trace anomaly constraint \eqref{T-constraints-summary}, the condition $\delta T = 0$ implies
\begin{equation}
    \delta R = 0 \, .
\end{equation}
While $R$ can be any constant over the solution space, we focus on the $R=0$ case. The  corresponding solution space has been constructed in \cite{Sheikh-Jabbari:2025kjd}:
\begin{equation}\label{q-T-conformal-infty}
    \begin{split}
        q_{ab}\, \d x^{a} \d x^{b} =& - h_+(x^+) h_-(x^-)\, \d x^{+} \d x^{-} \, , \\
        T_{\pm\pm} = - \frac{c}{12\pi \ell^2} h_{\pm}^2 &\, , \qquad
        T_{+-} = 0 \, ,\qquad \ttbarb=0,
    \end{split}
\end{equation}
where $ h_+(x^+)$ and $h_-(x^-)$ are arbitrary functions of their arguments which parametrize the conformal solution space.
The associated fluid variables take the form
\begin{equation}\label{asymp-conf-fluid-var}
    \rho = p = - \frac{c}{6\pi \ell^2} \, , \qquad
    u = \frac{h_+}{2}\,\partial_+ + \frac{h_-}{2}\,\partial_- \, .
\end{equation}
Noting that $\rho, p$ are constants and have no fluctuations, the phase and group velocities then follow immediately,
\begin{equation}
    v_p^2 = \frac{p}{\rho} = 1 \,, \qquad v_g=0\, .
\end{equation}
Thus, within the conformal solution space, while the asymptotic phase velocity is speed of light, the group velocity vanishes, which is a sign of having a null fluid discussed above. 

A comparison with the Dirichlet fluid is instructive. In the Dirichlet case, the dynamics is solely carried in the fluid energy-momentum  while the background geometry is fixed; as we have in the standard hydrodynamics. In contrast, in the conformal case, the fluctuations of the solution space are fully parametrized by $\delta h_+$ and $\delta h_-$. These fluctuations induce variations in the underlying metric, since $\delta \sqrt{-q} \neq 0$. Moreover, as follows from \eqref{asymp-conf-fluid-var}, the fluid velocity $u$ also fluctuates. Consequently, in the conformal case, the fluctuations of the induced metric and those of the boundary energy--momentum tensor are intertwined and fluctuate coherently. In other words, the fluid dynamics takes place on a background geometry that is itself partially dynamical.

\paragraph{Finite cutoff fluid.}
The fluid with asymptotic conformal boundary condition, at a finite radius $r$ is described by,
\begin{equation}\label{conformal-bc-flow}
\delta \left(\frac{\gamma_{ab}}{\sqrt{-\gamma}} \right) = - \frac{12\pi \ell^4}{c r^2}  \frac{r^{4}-\ell^4}{r^4+3 \ell^4}  \delta \left(\frac{\mathcal{T}_{ab}}{\sqrt{-\gamma}} \right), \, \qquad \delta {\cal T}=0\, ,
\end{equation}
which is a generic mixed boundary condition. 
In this case, the pressure and energy density are 
\begin{equation}
    p(r) = -\frac{c}{6\pi \ell^2}\frac{1}{1-\ell^2/r^2}\, , \qquad \rho(r) = -\frac{c}{6\pi \ell^2}\frac{1}{1+\ell^2/r^2}\, .
\end{equation}
From these expressions, we note that $\delta p(r) =0$ and hence $v_g(r)=0$.
\subsection{Asymptotic Neumann fluid}\label{sec:NBC}
In this subsection, we consider the asymptotic fluid with the Neumann boundary condition 
\begin{equation}
    \delta \left( \sqrt{-q} \, T^{ab} \right) = 0\, .
\end{equation}
Without loss of generality, we may choose \cite{Sheikh-Jabbari:2025kjd}  
\begin{equation}\label{N-bc-Sigma}
    \sqrt{-q}\, T^{ab} = \frac{c}{12 \pi \ell^2} 
    \begin{bmatrix}
        \Delta_1 & \Delta_2 \\
        \Delta_2 & 0
    \end{bmatrix}\,,
\end{equation}
in the $(t,\phi)$ basis where $\Delta_1$ and $\Delta_2$ are two constants with zero variations, $\delta \Delta_1=0=\delta \Delta_2$. The metric $q_{ab}$, takes the form  
\begin{equation}
    \mathrm{d}s_{q}^2 = q_{ab} \, \mathrm{d}x^{a} \mathrm{d}x^b = {2} F(t,\phi) \, \mathrm{d}t \, \mathrm{d}\phi + \frac{\Delta_1}{\Delta_2}[G(\phi)- F(t,\phi)] \, \mathrm{d}\phi^2 \,,
\end{equation}
with
\begin{equation}
    F(t,\phi) = -\frac{2\ell^2}{\Delta_1} \frac{\partial_{\phi}J(\phi) \partial_{\phi}H(\phi-\frac{2\Delta_2}{\Delta_1}t)}{[J(\phi)+H(\phi-\frac{2\Delta_2}{\Delta_1}t)]^2}\, ,
\end{equation}
where $G(\phi)$, $J(\phi)$ and $H(\phi-\frac{2\Delta_2}{\Delta_1}t)$ are arbitrary functions of their arguments. {Requiring that the solution has a smooth flat $\Delta_2\to 0$ limit yields,
\begin{equation}
    G(\phi)=F(t=0,\phi)= -\frac{2\ell^2}{\Delta_1}\frac{\partial_{\phi}J(\phi) \partial_{\phi}H(\phi)}{\left(J(\phi)+H(\phi)\right)^2}\, .
\end{equation}
So, the solution has two independent functions $J(\phi)$ and $H(\phi-\frac{2\Delta_2}{\Delta_1}t)$.}
The trace, $\ttbarb$, and the Ricci scalar are given as follows
\begin{equation}
    T = \frac{c\, \Delta_2}{6\pi \ell^2}\, , \qquad \ttbarb = - \frac{c^2 \Delta_2^2}{72 \pi^2 \ell^4} \, , \qquad R= -\frac{4\Delta_2}{\ell^2}\, .
\end{equation}
The Neumann fluid is described by a pure null fluid discussed in subsection \ref{subsec: asymp-fluid}, especially  \eqref{null-fluid-T}
\begin{equation}
    T_{ab} =  {\cal E}\, l_a l_b+\frac{c}{24\pi}  (l_a n_b+  n_a l_b) R\, .
\end{equation}
where
\begin{equation}
    \mathcal{E} = \frac{c\,\Delta_1\, G(\phi)}{3\pi \ell^2} \,, \qquad l_{a} \mathrm{d}x^a = -\frac{1}{2}\mathrm{d}\phi \,, \qquad n_{a} \d{}x^a = F(t,\phi) \d{}t + \frac{\Delta_1}{2\Delta_2}[G(\phi)- F(t,\phi)]\d{}\phi\, .
\end{equation}
As the above shows, for the Neumann case, $\delta R=0$ and $v_g=1$, as expected from a null fluid.

{For the flat $R=0$ case, which corresponds to $\Delta_2=0$ in our notation above,  $T, \ttbarb$ vanish and $T_{ab}={\cal E} l_a l_b$ \cite{Sheikh-Jabbari:2025kjd}. This case may be obtained as $\Delta_2\to 0$ limit in the above equations, yielding \cite{Sheikh-Jabbari:2025kjd}:
\begin{equation}
     \mathrm{d}s_{q}^2  = 2G(\phi) \, \mathrm{d}t \, \mathrm{d}\phi + t\ h(\phi)\, \mathrm{d}\phi^2 \,,
\end{equation}}
where
\begin{equation}
    h(\phi) : = -\frac{4\ell^2\, \partial_{\phi}J(\phi) }{\Delta_1\big( H(\phi) + J(\phi) \big)^3}\left[-2\big(\partial_{\phi}H(\phi)\big)^2 + \big(H(\phi)+J(\phi)\big)\partial_{\phi}^2H(\phi)\right]\, .
\end{equation}

The Neumann fluid may seem somewhat counterintuitive because we usually think of a fluid as a dynamical system evolving on a fixed background. In the Neumann case, however, the fluid itself is fixed, $\delta \left( \sqrt{-q} \, T^{ab} \right) = 0$, and all the dynamics are encoded in the induced metric, $\delta q_{ab} \neq 0$. In other words, as we move through the solution space, the fluid description remains unchanged, and only the underlying metric on which the fluid lives varies. 

\paragraph{Finite cutoff fluid.}
Now we discuss the finite cutoff fluid. We have
\begin{equation}
    \mathcal{T} = \frac{c \Delta_2}{6\pi \ell^2 + \frac{3\pi \ell^4}{r^2} \Delta_2}\, , \qquad \Ottbar = -\frac{c^2 \Delta_2^2}{72\pi^2 \ell^4\big(1 + \frac{\Delta_2}{2r^2}\big)^2}\, .
\end{equation}
The crucial feature of the Neumann boundary condition is that $v_g(r)=1$ and
\begin{equation}
    \mathcal{T}_{ab} =  {\cal E}(r)\, l_a(r) l_b(r) - \big(l_a(r) n_b(r)+  n_a(r) l_b(r) \big) \mathcal{T}  \, ,
\end{equation}
where
\begin{equation}
    \begin{split}
      &  {\cal E}(r) = \frac{ c \Delta_1 G(\phi)}{3\pi \ell^2 \Big(1 + \frac{\ell^2 \Delta_2}{2r^2}\Big)^4}\, , \\
      & l_{a}(r) \d{}x^a = - \frac{1}{2}\Big(1 + \frac{\ell^2 \Delta_2}{2r^2}\Big)^2 \d{}\phi\, , \\
      & n_{a}(r) \d{}x^a = F(t,\phi) \left[ \d{}t + \frac{\Delta_1}{2\Delta_2} \left( -1 + \frac{1 - \frac{\ell^2 \Delta_2}{2r^2}}{1 + \frac{\ell^2 \Delta_2}{2r^2}} \frac{G(\phi)}{F(t,\phi)} \right) \d{}\phi\right] \, ,
    \end{split}
\end{equation}
where $l(r)^2=0=n(r)^2$ and $l(r)\cdot n(r)=-1/2$.
\subsection{Asymptotic CSS fluid}
As the last example, we consider the fluid description with non-covariant boundary conditions \cite{Sheikh-Jabbari:2025kjd}, Compere-Song-Strominger (CSS) boundary conditions \cite{Compere:2013bya}
\begin{equation}
    \begin{split}
    &{q_{ab}} \d{}x^{a} \d{}x^{b}  =  2\d{}t \d{}\phi + J(\phi) \d{}\phi^2 \, ,\\
   &{T}_{ab} \d{}x^{a} \d{}x^{b} =  \Delta \d{}t^2+ \Delta J(\phi) \d{}t \d{}\phi  + [L(\phi)+\Delta J(\phi)^2]\d{}\phi^2\, ,
    \end{split}
\end{equation}
where $\phi\equiv \phi+2\pi$, $\Delta$ is a constant, and $J(\phi), L(\phi)$ are taken to be $2\pi$ periodic arbitrary functions.

The fluid variables associated with this boundary condition are
\begin{equation}
   \begin{split}
       & \rho=p= \frac{1}{2}[\Delta(4L+3\Delta J^2)]^{1/2}\, , \\
       & u^{a}\partial_{a} = \left(\frac{\Delta}{ 4L+3\Delta J^2 }\right)^{1/4} \left[ \partial_{\phi} -\frac{1}{2}\left( J+\sqrt{3J^2+4 \Delta^{-1}L} \right)\partial_t \right].
   \end{split}
\end{equation}
For this fluid, we also have
\begin{equation}
    v_{g}=1\, , \qquad v_p =1\, .
\end{equation}
As in the previous examples, one can study this case at a generic $r$, and find that $v_g(r)=1$ and 
\begin{equation}
    v_{p}^2(r) = \frac{1- \frac{6\pi \ell^4}{c r^2} \rho}{1 + \frac{6\pi \ell^4}{c r^2}\rho} = \frac{1- \frac{3\pi \ell^4}{c r^2} [\Delta(4L+3\Delta J^2)]^{1/2}}{1 + \frac{3\pi \ell^4}{c r^2}[\Delta(4L+3\Delta J^2)]^{1/2}}\, .
\end{equation}
As it is evident from the above expression, $v_{p}(r)\leq 1$. In a straightforward way and as in previous examples, one may work through the finite cutoff fluid at a given $r$, which we skip  here, as it does not add further specific information.

\section{Outlook}\label{sec:discussion}
As the next step in developing the freelance holography program started in \cite{Parvizi:2025shq, Parvizi:2025wsg} and as the direct continuation of the $3d$ freelance holography of \cite{Sheikh-Jabbari:2025kjd}, we have established the framework of \textit{freelance fluid/gravity correspondence} within AdS$_3$ gravity. By relaxing the standard holographic constraints, we constructed a hydrodynamic description applicable to any timelike boundary subject to arbitrary boundary conditions. In particular, we focused on the radial evolution of the system, interpreted as the Renormalization Group (RG) flow of the asymptotic fluid. We explicitly derived the flow of boundary conditions—physically corresponding to the flow among different hydrodynamic ensembles—and established an exact map interpolating the solution phase space between the asymptotic boundary and finite radial cutoffs. Furthermore, we conducted a detailed investigation into the RG flow of the fluid description itself, characterizing how the equation of state, energy conditions, and group velocities evolve along the radial direction, and in particular proved the $v_g$-theorem: the group velocity of sound waves is a monotonically decreasing as we move from the UV to the IR; this theorem parallels the famous $c$-theorem in 2D CFT's.

In this work, we focused on the AdS$_3$ gravity setup, where the equations of motion are fully integrable, allowing for exact control over the fluid dynamics. The extension of this framework to higher dimensions presents an interesting direction for future research, albeit with significant technical challenges. A fundamental distinction lies in the degrees of freedom: unlike the three-dimensional case, where the absence of bulk propagating modes results in a non-dissipative, integrable fluid, higher-dimensional gravity admits bulk gravitons that manifest as dissipative effects in the boundary hydrodynamics \cite{Adami:2021kvx, Ciambelli:2024kre, Arenas-Henriquez:2025rpt}. Consequently, higher-dimensional systems lack the exact radial integrability we exploited here. Propagating bulk modes effectively sources the radial evolution, necessitating the use of perturbative radial expansions rather than exact integration. While our 3D analysis characterized the RG flow of perfect fluid variables, a higher-dimensional treatment must essentially track the flow of transport coefficients, such as viscosity. Furthermore, higher dimensions exhibit the emergence of truly dynamical gravity at the boundary \cite{Adami:2025pqr}. In the AdS$_3$ case, while the finite-distance metric fluctuates, the boundary effective action lacks a standard kinetic term \cite{Sheikh-Jabbari:2025kjd}. In contrast, as demonstrated in \cite{Adami:2025pqr}, higher-dimensional theories naturally generate a kinetic term for the metric, yielding a fully dynamical fluid-plus-gravity system at finite distance. A detailed investigation of this rich hydro-gravitational dynamics is left for future work.

\begin{acknowledgments}
We thank M.H. Vahidinia for valuable discussions. We acknowledge the scientific atmosphere of the Simons Center workshop \textit{Timelike Boundaries in Classical and Quantum Gravity}, December 2025.  MMShJ gratefully acknowledges the hospitality of the Beijing Institute of Mathematical Sciences and Applications (BIMSA), where this project was initiated. The work of VT is supported by the Iran National Science Foundation (INSF) under project No. 4040771. The work of MMShJ is supported in part by the INSF Research Chair grant No. 4045163.
\end{acknowledgments}

\appendix

\section{Some useful relations}\label{appen:useful-rel}
Having integrated the $r$-dependence of Einstein's equations, one can work out useful identities relating finite arbitrary $r$ quantities to those at asymptotic large $r$:
\begin{subequations}\label{h-q--q-h}
\begin{align}
\gamma_{ab} = {\cal Q}_{ac} q^{cd}{\cal Q}_{db}\, ,&\qquad  {\cal Q}_{ab}:=q_{ab} - \frac{6\pi\ell^4}{c r^2}\tilde{T}_{ab}\, ,  
\label{hab-again}\\
q_{ab} =  {\cal H}_{ac} \gamma^{cd}{\cal H}_{db}\, ,& \qquad  {\cal H}_{ab}:=\gamma_{ab} +\frac{6\pi\ell^4}{c r^2}\tilde{{\cal T}}_{ab}\, , 
\label{qab-again}
\end{align}\end{subequations}
\begin{subequations}\label{All-r-quantities-1}
\begin{align}
({\cal Q}^{-1})^{ab}=   \frac{\sqrt{-q}}{\sqrt{-\gamma}}\ \Big(q^{ab}+\frac{6\pi\ell^4}{cr^2} T^{ab}\Big)\,, &\qquad ({\cal H}^{-1})^{ab} =  \frac{\sqrt{-\gamma}}{\sqrt{-q}}\  \Big(\gamma^{ab}-\frac{6\pi\ell^4}{c r^2} {\cal T}^{ab}\Big)\,, \label{Q-H-inverse}\\
{\cal Q}_{ab} = {\cal H}_{ab}\, ,&\qquad ({\cal Q}^{-1})^{ab} =  ({\cal H}^{-1})^{ab}\,,\label{Q-H}\\
       \tilde{\mathcal{T}}_{ab}  = {\cal Q}_{ac} q^{cd} \tilde{T}_{db} \, ,&\qquad 
     \tilde{T}_{ab} =  {\cal H}_{ac} \gamma^{cd} \tilde{\mathcal{T}}_{db}\,, \label{T-calT}\\
       \sqrt{-\gamma}\, \mathcal{T}^{ab} & = \sqrt{-q}\, (\mathcal{Q}^{-1})^{ac}q_{cd} T^{db}\, ,
\end{align}\end{subequations}
\begin{subequations}\label{All-r-quantities-2}
\begin{align}
   \sqrt{-\gamma}\, \mathcal{T} = \sqrt{-q} \Big(T-\frac{6\pi \ell^4}{c r^2} \ttbarb \Big)\, ,& \qquad 
   \sqrt{-q}\, {T} = \sqrt{-\gamma} \Big({\cal T}+\frac{6\pi \ell^4}{c r^2} \Ottbar\Big)\, ,\label{qT-hcalT}\\ 
    \sqrt{-\gamma}\, {\cal R} =\sqrt{-q}\, {R}\, &,\qquad
    \sqrt{-\gamma}\, \Ottbar = \sqrt{-q}\ \ttbarb\,, \label{r-dep-ttbar-3}\\
      \sqrt{-\gamma} = \sqrt{-q} \left( 1 -\frac{\ell^4}{4r^2} {R}  -\frac{18 \pi^2 \ell^8 \ttbarb}{c^2 r^4}\right)\, &,\qquad    \sqrt{-q} =  \sqrt{-\gamma} \left( 1+ \frac{\ell^2}{4} {\cal R}  +\frac{18 \pi^2 \ell^4}{c^2 }\Ottbar\right)\, ,\label{detq-deth} 
\end{align}\end{subequations}
where $\ttbarb$ is the T$\bar{\text{T}}$ operator associated with boundary EMT
\begin{equation}
    \ttbarb := T^{ab} T_{ab} - T^2\, .
\end{equation}
In deriving the above equations, we have used the following $2d$ identities
\begin{equation}
    \begin{split}
        T_{a}^{c} T_{cb} = T T_{ab} + \frac{1}{2} \ttbarb q_{ab}\, ,\qquad \tilde{T}_{a}^{c} \tilde{T}_{cb} = -T \tilde{T}_{ab} + \frac{1}{2} \ttbarb q_{ab}\, , \qquad {\tilde T}_{a}^{c} T_{cb} = \frac{1}{2} \ttbarb q_{ab}\, .
    \end{split}
\end{equation}

\bibliographystyle{fullsort.bst}
\bibliography{reference}
\end{document}